\newcommand{\be}{\begin{equation}}
\newcommand{\ee}{\end{equation}}

\documentclass[osa,osajnl,twocolumn,amsmath,amssymb]{revtex4}

\usepackage{graphicx}
\usepackage{dcolumn}
\usepackage{bm}
\usepackage{amsmath}

\begin{document}
\title{Stable rotating dipole solitons in nonlocal optical media}

\author{Servando Lopez-Aguayo, Anton S. Desyatnikov, and Yuri S. Kivshar}

\affiliation{Nonlinear Physics Center, Research School of Physical Sciences and Engineering, Australian National University, Canberra, ACT 0200, Australia}

\author{Stefan Skupin and Wieslaw Krolikowski}

\affiliation{Laser Physics Center, Research School of Physical Sciences and Engineering, Australian National University, Canberra, ACT 0200, Australia}

\author{Ole Bang}

\affiliation{COM$\bullet$DTU Department of Communications, Optics \& Materials, Technical University of Denmark, DK-2800 Kongens Lyngby, Denmark}

\begin{abstract}
We reveal that nonlocality can provide a simple physical mechanism for stabilization of multi-hump optical solitons, and present the first example of stable rotating dipole solitons and soliton spiraling, known to be unstable in all types of realistic nonlinear media with local response.
\end{abstract}


\maketitle

Recently increased interest in the study of self-trapped optical beams in nonlocal nonlinear media is explained by experimental observations of nonlocal spatial solitons in liquid crystals~\cite{Assanto} and lead glasses~\cite{moti} as well as a number of many interesting theoretical predictions~\cite{W1,W2,John,Bang}, including stabilization of vortex solitons against symmetry breaking instability in the media with nonlocal optical response~\cite{vort1,vort2}. Many of the predicted and demonstrated properties of nonlocal nonlinear modes suggest that in such optical media we should expect stabilization of many different types of spatial nonlinear structures such as necklaces~\cite{neck}, soliton clusters~\cite{cluster}, and recently introduced broader classes of modulated optical vortices or azimuthons~\cite{azimuthon}.

One of the simplest multi-hump solitons predicted in nonlinear optics is a dipole-like structure composed of two interacting fundamental beams with opposite phases that undergo angular rotation during the propagation~\cite{book,PIO}. Such a localized structure can be viewed as a special type of strongly modulated single-charge vortex soliton or azimuthon and it can also be linked to the problem of soliton spiraling, as discussed e.g. in \S 6.1 of Ref.~[13]. However, in spite of its broad applicability and deep physical context, the rotating dipoles are known to be {\em always unstable} in all types of realistic optical media with local response~\cite{book,PIO}.

In this Letter, we study the effect of nonlocal response of an optical medium on the stability of rotating dipole solitons. We demonstrate that such solitons can be stabilized when the nonlocality parameter exceeds some threshold value that decreases with the angular momentum and angular velocity, i.e. the dipoles rotating faster are more robust. We describe a continuous family of two-lobe localized structures (or azimuthons~\cite{azimuthon}) that extend from non-rotating dipole solitons to radially symmetric vortex solitons by continuous azimuthal transformations.

We consider the propagation of paraxial optical beams in a nonlinear medium described by the nonlinear Schr\"{o}dinger equation for the scalar field envelope $E$~\cite{book}
\be \label{NLS} i\partial_zE+\nabla^2 E+n\left(I,{\vec r}\right)E=0, \ee
where $z$ and $\vec{r}$ stand for the propagation and transverse coordinates, respectively. We assume that the nonlinear refractive index $n$ depends on the intensity $I\equiv|E|^2$ via the following nonlocal relation
\be\label{nonlocality}
n(I,{\vec r})=\int R\left(|{\vec r}-{\vec \rho}|\right)I\left({\vec \rho}\right)d{\vec\rho},
\ee
where the response function $R(r)$ is determined by the specific physical process responsible for the medium nonlinearity. Here we consider a Gaussian nonlocal response %%
\be\label{R}
R(r)=\sigma^{-2}\pi^{-1}\exp\left(-r^2/\sigma^2\right),
\ee
where $\sigma$ measures the degree of nonlocality. In the following we use the scaling ${\vec r}={\vec r}\,'\sigma$, $z=z'\sigma^2$, $E=E'/\sigma$, and omit primes.

\begin{figure}
\includegraphics[width=84mm]{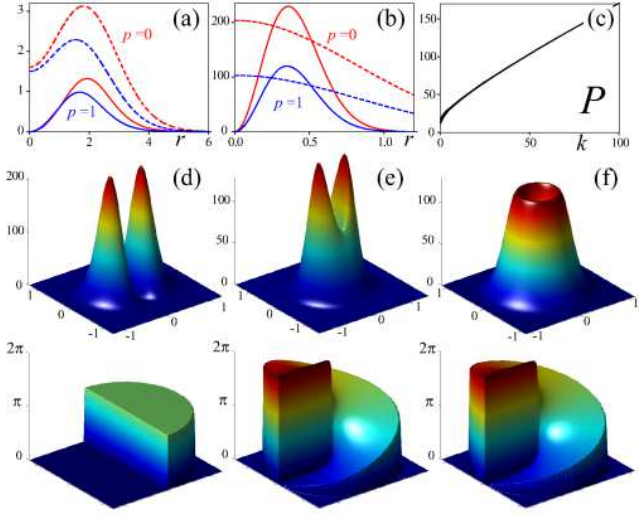}%\vspace{-2mm}
\caption{\label{fig1} (Color online) Beam radial intensity $U^2(r)$ (solid) and effective nonlinear potential $N(U,r)$ (dashed) for the dipole soliton ($p=0$, red) and vortex soliton ($p=1$, blue) for (a) relatively low nonlocality with $k=1$ and (b) high nonlocality with $k=70$. (c) Power vs. propagation constant $k$ for the dipole solitons; several lines for different $p$ coincide. (d)-(f) Envelopes of the intensity (top) and phase (bottom) for three characteristic cases with $k=70$ and (d) $p=0$, (e) $p=2^{-1/2}$, and (f) $p=1$.}
\end{figure}

Multi-hump localized solutions of this model can be found numerically by two-dimensional relaxation methods~\cite{Mironov}. However, this approach represents a major difficulty for the complex envelopes with singular phase profiles due to a poor convergence. Therefore, here we employ an approximate but much simpler variational method~\cite{var} allowing a reasonably good approximation~\cite{azimuthon}. To this end, we introduce the following ansatz
\be\label{ansatzp} E=\sqrt\pi\,U\left(r\right)\left[\cos(m\varphi)+i
p \sin(m\varphi)\right]\exp\left(ikz\right), \ee
where $\varphi$ is the azimuthal angle and the parameter $p$ ($0 \leq p \leq 1$) determines the modulation depth (contrast) of the beam intensity, $n=1-p^2$. Substituting this ansatz into the averaged Lagrangian~\cite{var} associated with Eq.~(\ref{NLS}), we obtain the equation for the radial amplitude $U(r)$ of the localized mode,
\be\label{stat} -kU+U_{rr}+r^{-1}U_r-m^2 r^{-2}U + U N(U^2,r)=0,
 \ee
where the nonlinear term is given by the expression
\begin{eqnarray}
\nonumber N(U^2,r)&=&\pi(1+p^2)\,{\rm e}^{-r^2}\int_0^\infty \rho d\rho\,{\rm e}^{-\rho^2}\,U^2(\rho)\times \\
\label{convvar}&&\left\{I_0(2r\rho)
+\frac{1}{2}\left(\frac{1-p^2}{1+p^2}\right)^2
I_{2m}(2r\rho)\right\},
\end{eqnarray}
and $I_j$ is the modified Bessel function of the first kind. In the limit $p\to 1$, the function (\ref{ansatzp}) becomes radially symmetric, and the potential~(\ref{convvar}) corresponds to a vortex soliton. On the other hand, for $p\to 0$, this solution describes {\em scalar multipole solitons} (or optical necklaces) known to be radially unstable in local media.

Applying to Eq.~(\ref{stat}) the numerical shooting technique and an iteration procedure, we find the stationary localized solutions with different values of $p$ and $k$. Our results are summarized in Figs.~\ref{fig1}(a-f). Figures~\ref{fig1}(a,b) show the radial envelopes of the localized modes (solid lines) and the corresponding profiles of the effective refractive index (dashed lines) for the dipole ($p=0$) and vortex ($p=1$) solitons in the case of low [(a)] and high [(b)] degree of nonlocality. Figure~\ref{fig1}(c) shows the dependence of the soliton power $P=\pi(1+p^2) \int_0^\infty U^2rdr$ on the propagation constant $k$, for several values of the modulation parameter $p$. Figures~\ref{fig1}(d-f) present the intensity (top) and phase (bottom) distributions for the dipole, azimuthon, and vortex solitons, respectively.

Having a nontrivial phase structure, the azimuthons carry {\em a nonzero angular momentum}. Using the ansatz Eq.~(\ref{ansatzp}), we find that the beam orbital angular momentum, $M={\rm Im}\int E^*\partial_\varphi E\,d\vec{r}$, normalized to its power $P$, defines the beam effective (fractional) spin, $S\equiv M/P=2p/(1+p^2)$.

These results show the effect of the modulation depth (determined by $p$) on the soliton intensity, and they demonstrate that a transition from a dipole mode to a radially symmetric vortex soliton can be achieved by continuous azimuthal transformations. Therefore, the two-peak azimuthon solutions described above provide a link between two spiraling solitons with opposite phase and radially symmetric optical vortices, or spatially localized singular beams.

\begin{figure}
\includegraphics[width=84mm]{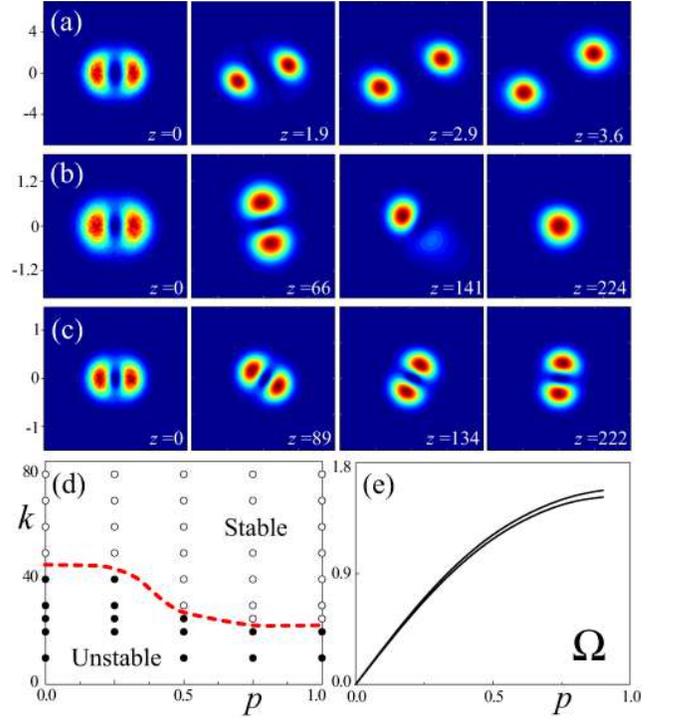}%\vspace{-2mm}
\caption{\label{fig2} (Color online) Two different instability scenarios for rotating dipole solitons with $p=0.5$ for (a) $k=1$ and (b) $k=20$. (c) Stable rotation for $p=0.5$ and $k=100$. In all three cases initial noise is 10\%. (d) Stability domain of dipole solitons in the ($p,k$) space. (e) Numerically found relation between the angular velocity $\Omega$ of a dipole soliton and its contrast parameter $p$.}
\end{figure}

In order to study stability of the stationary solutions we solve Eq.~(\ref{NLS}) numerically by the split-step beam propagation method with a fast Fourier transform. As initial conditions, we take the stationary solutions perturbed initially by $10-20\%$ of noise and propagate them over the distances of $z\sim 10^2-10^3$ or larger.

\begin{figure}
\includegraphics[width=84mm]{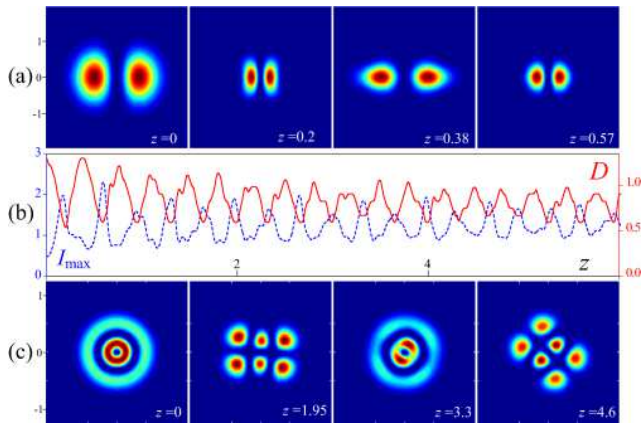}%\vspace{-2mm}
\caption{\label{fig3} (Color online) (a,b) Evolution of a dipole soliton ($k=60$, $p=0$) excited by a strong initial deformation. Persistent oscillations of both the mode amplitude $I_{\rm max}$ (dashed) and diameter $D$ (a distance between two peaks, solid) are observed; these oscillations decay slow during the propagation. (c) Example of the soliton revivals observed in the instability-induced evolution of double-ring single-charge vortex solitons with $k=100$.}
\end{figure}

Figure~\ref{fig2} presents typical propagation scenarios. When the nonlocality is weak  the dipole breaks into two mutually repelling filaments [(a)], and this behavior is reminiscent of the azimuthal instability of vortices in local media~\cite{PIO}. For the subcritical nonlocality [(b)] the azimuthon initially breaks up into two filaments which are subsequently forced by the nonlocality-induced potential to collide and merge into a single fundamental soliton. Stable propagation of the rotating azimuthon is observed when nonlocality exceeds a critical value [(c)]. This corresponds to a balance of the effective nonlocality-mediated attractive force and centrifugal repulsive force. The domain of stability on the plane $(p,k)$ is shown in Fig.~\ref{fig2}(d), and the stability threshold decreases in the limit of the radially symmetric vortices ($p\to 1$). For the stable rotating dipoles, we obtain the angular velocity $\Omega$ by numerical averaging and Fig.~\ref{fig2}(e) shows the dependence of the azimuthon angular velocity on the modulation parameter $p$. Velocity increases monotonically with $S$ and only weakly depends on the power (two curves are for $k=60$ and $k=80$). For the vortex soliton with $p=1$ the rotation velocity is a free parameter.

We have tested the stability of the rotating dipole solitons with respect to strong perturbations which lead to large-amplitude oscillations, as shown in Fig.~3(a,b). Similar studies have been carried out for the spatial solitons described by Eq.~(\ref{ansatzp}) with higher azimuthal indices $m>1$ (not shown). Somewhat surprising results we obtained by testing stability of double-ring radially symmetric ($p=1$) solitons with $m=1$. While no absolutely stable solitons have been found in our numerical simulations, we have observed the effect of soliton revivals after modulational instability is fully developed, as shown in Fig.~3(c). Such an instability can be easily excited if we perturb the stationary localized mode by a small amount $\delta$ as follows, $U(r)[\cos(m\varphi)+ i\sin(m\varphi+\delta)]$. The structures which appear at intermediate stages of this instability dynamics are similar to higher-order linear modes such as Laguerre- and Hermite-Gaussian modes. Thus, in this case the nonlocal medium acts as a self-induced mode converter, where the conversion dynamics depends on the initial mode and the properties of the medium, i.e. the value of the nonlocality parameter. After several revivals, the structure losses its symmetry and undergoes irregular dynamics.

In conclusion, we have demonstrated that nonlinear optical media with a nonlocal response can stabilize rotating dipole solitons and their generalizations in the form of modulated azimuthons with a fractional spin. The stabilization is achieved when the nonlocality parameter exceeds a certain threshold value, and faster rotating azimuthons are found to be more robust.

S. Lopez-Aguayo is with Photonics and Mathematical Optics Group, Tecnologico de Monterrey, M\'exico.

\end{document}